\documentclass{PoS}

\usepackage{bm}
\usepackage{url}

\newcommand{\ie}{\textit{i.e.}\ }
\newcommand{\eg}{\textit{e.g.}\ }
\newcommand{\cf}{\textit{c.f.}\ }

\newcommand{\ns}{n_{\mathrm{s}}}
\newcommand{\thetas}{\vec{\theta}_{\mathrm{s}}}

\newcommand{\pypackage}[1]{\textit{#1}}
\newcommand{\pymodule}[1]{\texttt{#1}}
\newcommand{\pyclass}[1]{\texttt{#1}}
\newcommand{\pymethod}[1]{\texttt{#1}}

\title{SkyLLH - A generalized Python-based tool for log-likelihood analyses in multi-messenger astronomy}

\ShortTitle{SkyLLH - A generalized Python-based tool for log-likelihood analyses}

\author{
The IceCube Collaboration\footnote{For collaboration list, see PoS(ICRC2019) 1177.}\\
{\itshape \href{http://icecube.wisc.edu/collaboration/authors/icrc19_icecube}{http://icecube.wisc.edu/collaboration/authors/icrc19\_icecube}}\\
E-mail: \email{martin.wolf@icecube.wisc.edu}
}

\abstract{
Common analysis techniques in multi-messenger astronomy involve hypothesis tests with unbinned log-likelihood (LLH) functions using data recorded in celestial coordinates to identify sources of high-energy cosmic particles in the Universe.
We present the new Python-based tool ''SkyLLH`` to develop such analyses in a telescope-independent framework. The main goal of the software is to provide an easy-to-use and modularized concept to implement and to execute such LLH functions efficiently on the computer with high-performance. SkyLLH can be applied on different multi-messenger data like neutrino and gamma-ray events from experiments such as the IceCube Neutrino Observatory and the Fermi-LAT. In this contribution we highlight SkyLLH's various design goals, current development status, and prospects for its wider application in multi-messenger astronomy.
\\

\vspace{4mm}
{\bfseries Corresponding authors:}
\speaker{Martin Wolf}$^{1}$\\
{$^{1}$ \itshape Physik-department, Technische Universit\"at M\"unchen, D-85748 Garching, Germany}

}

\FullConference{36th International Cosmic Ray Conference -ICRC2019-\\
		July 24th - August 1st, 2019\\
		Madison, WI, U.S.A.}

\begin{document}

\section{Introduction}\label{sec:intro}

In multi-messenger astronomy (MMA) the maximum log-likelihood (LLH) hypothesis ratio test (LRT) is a common analysis technique. Let $\Theta$ denote the entire allowed parameter space of the model with model parameters $\vec{\theta}$, the LRT is constituted of the LLH ratio test statistic, $\log \Lambda(\vec{D})$, for a null- and alternative hypothesis, $\mathcal{H}_0: \vec{\theta} \in \Theta_0$, and $\mathcal{H}_1: \vec{\theta} \in \Theta_0^c$, respectively:
\begin{equation}
    \log \Lambda(\vec{D}) = \sup_{\vec{\theta} \in \Theta_0} \left\{ \log \mathcal{L}(\vec{\theta}|\vec{D}) \right\} - \sup_{\vec{\theta} \in \Theta} \left\{ \log \mathcal{L}(\vec{\theta}|\vec{D}) \right\},
\label{eq:llh-ratio}
\end{equation}
where $\log \mathcal{L}(\vec{\theta}|\vec{D})$ is the logarithm of the likelihood function with parameter set $\vec{\theta}$ for the given (observed) data $\vec{D}$. In the limit of large statistics the probability density function (PDF) of the test-statistic $\lambda(\vec{D}) = -2\log\Lambda(\vec{D})$, \textit{i.e.} $f(\lambda)$, follows a $\chi^2_k$-distribution with the number of degrees of freedom, $k$, equal to the number of free parameters in the test, if the parameter values are sufficiently far away from their bounds \cite{Wilks:1938dza}.

In MMA a usual task is to analyze data composed of independent events, so-called event data, with contributions from a signal and a background component. Hence, a two-component, \textit{i.e.} signal and background, likelihood model is usually chosen.
In such a model a likelihood function $\mathcal{L}(\vec{\theta}|\vec{D})$ can be constructed, that evaluates the observable PDFs for each component, \ie signal and background, of each of the $N$ data events, $D_i$:
\begin{equation}
    \mathcal{L}(\vec{\theta}|\vec{D}) \equiv \mathcal{L}(\ns,\thetas|\vec{D}) = \prod_{i=1}^{N}\left[ \frac{\ns}{N} \mathcal{S}(D_i|\thetas) + (1 - \frac{\ns}{N}) \mathcal{B}(D_i) \right].
\label{eq:tc-likelihood-function}
\end{equation}
Here, the parameter set $\vec{\theta}$ consists of the mean number of signal events in the data, $\ns$, and the parameters $\thetas$ describing the signal source. $\mathcal{S}$ and $\mathcal{B}$ are the signal and background probability density functions for the $N$ data events $D_i$, respectively.

In the special case, when assuming a negligible signal contribution, \textit{e.g.} in a background-dominated experiment, the null-hypothesis can be defined as the data without any signal events, \textit{i.e.} $\mathcal{H}_0: n_{\mathrm{s},0}=0$. Thus, the negative of the log-likelihood ratio test statistic for the two-component likelihood function (\ref{eq:tc-likelihood-function}) can be written as
\begin{equation}
    -\log \Lambda(\vec{D}) = \sup_{\ns,\thetas} \left\{ \log \frac{\mathcal{L}(\ns,\thetas|\vec{D})}{ \mathcal{L}(\ns=n_{\mathrm{s},0}=0|\vec{D})} \right\} = \sup_{\ns,\thetas} \left\{ \sum_{i=1}^N \log \left[ \frac{\ns}{N}\left(\frac{\mathcal{S}(D_i|\thetas)}{\mathcal{B}(D_i)} - 1\right) + 1 \right] \right\}.
\end{equation}
It should be noted that the PDF $f(\lambda)$ of the test statistic function $\lambda(\vec{D})$ does not follow a $\chi^2$-distribution function under this null-hypothesis, because the parameter value for $\ns$ lies on the boundary of the allowed parameter value space $[0,N]$. Instead, $f(\lambda)$ follows the super-position of a $\delta$-function and a $\chi^2_k$-distribution function \cite{Cowan:2010js}:
\begin{equation}
    f(\lambda) = \frac{1}{2}\delta(\lambda) + \frac{1}{2}\chi^2_k(\lambda)   
\end{equation}

The SkyLLH software provides a framework for defining maximum LLH hypothesis ratio tests, to construct, and to evaluate LLH ratios as given in equation (\ref{eq:llh-ratio}). It allows the user to define a statistical LRT analysis for a given set of data. Furthermore, the user can easily generate pseudo data trials to determine statistical properties of the analysis, \eg sensitivity and discovery potential, in a frequentist approach.

\section{The SkyLLH Framework}

The SkyLLH framework is implemented entirely in the programming language Python\footnote{Both, Python2 and Python3 environments are supported. However, Python3 based developments of analyses, that utilize SkyLLH, is preferred.} and is an open-source package licensed under the GPLv3 license\footnote{\url{https://www.gnu.org/licenses/gpl-3.0.txt}}. It is available on the open source github repository\footnote{\url{https://github.com/IceCubeOpenSource/SkyLLH/}} of the IceCube collaboration. The software package dependencies for SkyLLH are kept to an absolute minimum: \pypackage{numpy}, \pypackage{scipy}, and \pypackage{astropy}. These packages are usually already available in most Python installations. In cases where they are not, they are at least easily available for installation on most platforms.

SkyLLH utilizes object-oriented-programming (OOP) techniques. The class structure is tied to the mathematical objects of the LLH ratio formalism. A fundamentally used mathematical object is the PDF. Thus, the Python abstract base class \pyclass{core.PDF} exists and defines the interface of a PDF object. The most important method of this class is \pymethod{PDF.get\_prob}, which is supposed to return the probability for each data event $D_i$.

The code of the SkyLLH framework is structured into four distinct main Python modules: core, physics, plotting, and detector specific implementations and specializations. The \pymodule{core} module contains all Python classes defining the base framework of SkyLLH. This includes the classes for all mathematical objects like PDFs, LLH ratio functions, and test statistic functions. The \pymodule{physics} module provides classes defining the physics models of interest. Since the actual sources of high-energy cosmic rays, including photon and neutrino emission, remain unknown to-date, commonly used physics models are generic flux models for the messenger particle of interest that creates a signal in the particular detector. The \pymodule{plotting} module contains auxiliary functionality for plotting particular objects of the framework, \eg PDF objects. This is meant mainly for integrity checks of the analysis. Finally, the fourth part of modules provide detector specific implementations and specializations based on the classes defined in the \pymodule{core} module. Currently this includes the \pymodule{i3} module for the IceCube Neutrino Observatory \cite{Aartsen:2016nxy}, which is a cubic-kilometer neutrino detector installed in the ice at the geographic South Pole between depths of 1450~m and 2450~m, and was completed in 2010.

\section{Source Hypothesis Definition}

The first step of a LRT in MMA is the definition of the source hypothesis. Because the actual sources of high-energy cosmic rays in the Universe are still unknown, a common approach is to define a generic differential particle flux at Earth, $\Phi_\mathrm{S}$, from a source. Such a generic differential flux can be parameterized as
\begin{equation}
    \frac{\mathrm{d}^4\Phi_\mathrm{S}(\alpha,\delta,E,t|\thetas)}{\mathrm{d}E\mathrm{d}\Omega\mathrm{d}A\mathrm{d}t},
    \label{eq:generic-diff-flux}
\end{equation}
which is a function of the celestial coordinates right-ascention, $\alpha$, and declination, $\delta$, as well as the energy and time of the source (signal) particle, given the source parameters $\thetas$. $\Omega$ and $A$ denote the sky's solid-angle covered by the source, and the surface area of Earth covered by the flux, respectively. $\Phi_{\mathrm{S}}$ can describe a point-like source, like a Blazar, or an extended source, like the galactic plane of the Milky Way.

The IceCube Neutrino Observatory searches for point-like sources using a factorized flux model as source hypothesis:
\begin{equation}
    \frac{\mathrm{d}^4\Phi_\mathrm{S}(\alpha,\delta,E,t|\thetas)}{\mathrm{d}E\mathrm{d}\Omega\mathrm{d}A\mathrm{d}t} = \Phi_0 \Psi_{\mathrm{S}}(\alpha,\delta|\thetas)\epsilon_{\mathrm{S}}(E|\thetas)T_{\mathrm{S}}(t|\thetas),
    \label{eq:factorized-flux-model}
\end{equation}
where $\Phi_0$ is the flux normalization carrying the differential flux units, and $\Psi_{\mathrm{S}}$, $\epsilon_{\mathrm{S}}$, and $T_{\mathrm{S}}$ are the spatial, energy, and time profiles of the source, respectively. For a point-like source at the celestial location $\vec{r}_{\mathrm{s}} = (\alpha_{\mathrm{s}},\delta_{\mathrm{s}})$ the spatial profile collapses to a Dirac-$\delta$-function:
\begin{equation}
    \Psi_{\mathrm{S}}(\vec{r}|\thetas) = \delta(\vec{r}-\vec{r}_{\mathrm{s}}),
\end{equation}
with $\vec{r} = (\alpha,\delta)$ being the celestial coordinate vector. 
As energy profile $\epsilon_{\mathrm{S}}(E|\thetas)$ a common choice is the power-law function
\begin{equation}
    \epsilon_{\mathrm{S}}(E|\thetas) = \left(\frac{E}{E_0}\right)^{-\gamma}
    \label{eq:power-law-profile}
\end{equation}
with spectrial index $\gamma$ and reference energy $E_0$. The time profile $T_{\mathrm{S}}(t|\thetas)$ is usually choosen to be box-shaped or gaussian-shaped. For a steady emitting source the time profile is unity.

SkyLLH provides the abstract base class \pyclass{physics.flux\_models.FluxModel} for the generic differential flux given by (\ref{eq:generic-diff-flux}). Utilizing the \pymodule{units} module of the \pypackage{astropy} package, this class supports the specification of individual energy, angle, length, and time units for the flux, as well as the their conversion to the internally used flux unit of GeV$^{-1}$~sr$^{-1}$~cm$^{-2}$~s$^{-1}$. In addition \pyclass{FactorizedFluxModel} provides a class for a factorized flux model as given in equation (\ref{eq:factorized-flux-model}). Individual spatial, energy, and time flux profiles can be designed through the abstract base class \pyclass{FluxProfile}. 

In case the LRT consists of several sources, it is useful, from a computational point-of-view, to group sources with the same flux model into source hypothesis groups. Each group of sources will share the same implementation for the flux model, the mean number of expected signal events in the detector, \ie the detector signal yield, and the signal event generation method. Calculations for those quantities can then be parallelized for all sources of a source hypothesis group. SkyLLH provides the \pyclass{SourceHypoGroup} and the \pyclass{SourceHypoGroupManager} classes to define and to mange those groups of source hypotheses.

\section{Data Management Concepts}

An important utility feature of SkyLLH is the ability to pre-define common data sets that can be used by different analyses. To accommodate this feature, the \pymodule{core.dataset} module provides the \pyclass{Dataset} class. An instance of the \pyclass{Dataset} class describes a particular data set. The general enfolded properties of a data set are its name, the location on disk of the experimental and simulation data, as well as its live-time. Data can be stored on disk in various data formats. Depending on the file name extension of the data files, SkyLLH selects a specific file loader class for a particular data file format. The list of file loader classes is extendable by the user in order to be able to support user-specific data file formats. Sometimes a data sample is split into several individual data sets. For instance when an event selection has to be performed on individual detector configurations or calibrations, resulting into separate detector responses for each data set. In SkyLLH such individual data sets can be grouped into a collection of data sets by using an instance of the \pyclass{core.dataset.DatasetCollection} class. 
In addition to the data set properties listed above, the \pyclass{Dataset} class supports the version and version qualifiers properties to allow for a documented evolution of data sets. This simplifies the reproduce-ability of the results of analyses.

After the pre-definition of data sets and collection of data sets, \ie data samples, such data sets can be chosen by an analysis. For each data set the experimental, simulation, and possible auxiliary data can be loaded with a single call to the \pymethod{load\_and\_prepare\_data} method of the \pyclass{Dataset} class. This class method applies possible data preparation functions to achieve uniformity in available data fields throughout the data sets. These data preparation functions also allow for analysis related data selection cuts.
The loaded data is then stored in an instance of the data holder class \pyclass{DatasetData} and can be used for creating PDF objects, background and signal event generator instances. 

Internally, the data is stored as \pypackage{numpy} arrays. A dedicated class named \pyclass{core.storage. DataFieldArray} has been developed for SkyLLH to store the data fields of a data set as one-dimensional \pyclass{numpy.ndarray} objects, whereas the interface of this class mimics the one of a structured ndarray object. Because numerical operations for the analysis are usually performed on a single data field for all events of the data set at once, data access is more efficient on one-dimensional ndarrays, where field data of consecutive events are stored continuously in memory, compared to structured ndarrays where entire event records are stored continuously in memory, resulting into a much larger memory footprint when traversing a single data field for all events. By using \pyclass{DataFieldArray} as internal data holder, memory cache misses can be reduced, causing faster data access and shorter program execution times.

Another unique data management concept of SkyLLH is the management of trial data.
For each analysis trial new pseudo event data has to get generated, which is based, in some analysis defined way, on the loaded data from the pre-defined data sets. The likelihood function $\mathcal{L}(\vec{\theta}|\vec{D})$ is always evaluated on the trial data. In the case of data unblinding, the trail data equals the experimental data. In SkyLLH the \pyclass{core.trialdata.TrialDataManager} class provides a manager for the event data of a trial. The manager allows to define additional event data fields and their calculation based on the data fields of the data set and previously defined data fields for the trial data manager. \pyclass{TrialDataManager} data fields have assigned possible dependencies on source properties and fit parameters. Hence, only those data fields have to be recalculated, whose dependencies were updated during the LLH function maximization process and the data $\vec{D}$ for the LLH function $\log \mathcal{L}(\vec{\theta}|\vec{D})$ is requested solely through the \pyclass{TrialDataManager} instance of the data set. 

\section{Hypothesis Parameter Definition}

For the LRT the definition of the parameter set $\Theta$ is essential. In general it can contain fixed and floating, \ie fit, parameters $\vec{\theta}$. For the signal-and-background two-component likelihood model, a parameter $\theta$ belongs either to a signal, a background, or some other model, for instance a nuisance model. We denote the parameter $\theta$ as a global parameter and define a local parameter, $\tilde{\theta}$, as a parameter of a specific model. In general the signal component can consist of multiple contributions, \eg multiple individual point-like neutrino emitting sources in the Universe. Each signal model would represent a single source, modeled for instance by a neutrino flux as given in equation (\ref{eq:factorized-flux-model}). Such a scenario is commonly referred to as "stacking of sources``.

A specific model has a distinct set of parameters with their names. Hence, in the special case of a multi point-like source scenario the flux model, as given in equation (\ref{eq:factorized-flux-model}) with the energy profile of a power-law (\cf eq. (\ref{eq:power-law-profile})), has the spectral index (local) parameter $\tilde{\gamma}$, named "gamma", defined several times, once for each source.
It raises the question how these local parameters should be handled in the global likelihood maximization process. Obviously, this question is of hypothesis nature. Depending on the source hypothesis these local spectral index parameters should all be independent global fit parameters, or should all share the same single global spectral index parameter, or should be grouped into groups of shared global spectral index parameters. SkyLLH provides the \pyclass{ModelParameterMapper} class that takes a global parameter and assigns it to a local parameter of one or more models. Hence, global fit parameters can be assigned to local model parameters in a highly flexible manner.

\section{Application of SkyLLH in Multi-Messenger Astronomy}

SkyLLH is being developed within the IceCube collaboration as a standard tool to search for neutrino emitting sources in the Universe. The implementation of generalized concepts in terms of source hypothesis and hypothesis parameter definition makes it easy to use the SkyLLH framework also for searches of other messenger particles in other experiments. Whenever a LRT as given in equation (\ref{eq:llh-ratio}) with celestial data has to be performed, SkyLLH is a suitable tool. Possible future applications of SkyLLH could be combined analyses of same-kind messenger particle data, for instance from different neutrino telescopes like IceCube and ANTARES~\cite{Collaboration:2011nsa} / KM3NeT~\cite{Adrian-Martinez:2016fdl}, or of different messenger particle data of neutrinos and gamma-rays, for instance from IceCube and Fermi/LAT~\cite{Atwood:2009FermiLAT}.

%


\bibliographystyle{ICRC}
\bibliography{main}

\providecommand{\href}[2]{#2}\begingroup\raggedright\begin{thebibliography}{1}

\bibitem{Wilks:1938dza}
S.~S. Wilks, {\em Annals Math. Statist.} {\bf 9} (1938) 60--62.

\bibitem{Cowan:2010js}
G.~Cowan, K.~Cranmer, E.~Gross, and O.~Vitells, {\em Eur. Phys. J.} {\bf C71}
  (2011) 1554. [Erratum: Eur. Phys. J.C73,2501(2013)].

\bibitem{Aartsen:2016nxy}
{\bf IceCube} Collaboration, M.~G. Aartsen et~al., {\em JINST} {\bf 12} (2017)
  P03012.

\bibitem{Collaboration:2011nsa}
{\bf ANTARES} Collaboration, M.~Ageron et~al., {\em Nucl. Instrum. Meth.} {\bf
  A656} (2011) 11.

\bibitem{Adrian-Martinez:2016fdl}
{\bf KM3Net} Collaboration, S.~Adrian-Martinez et~al., {\em J. Phys.} {\bf G43}
  (2016) 084001.

\bibitem{Atwood:2009FermiLAT}
{\bf Fermi/LAT} Collaboration, W.~B. Atwood et~al., {\em Astrophys. J.} {\bf
  697} (2009) 1071.

\end{thebibliography}\endgroup

\end{document}